\documentclass[%
 aip,
a4paper,
amsmath,amssymb,preprint,%
]{revtex4-1}

\usepackage{graphicx}
\usepackage{dcolumn}
\usepackage{bm}

\usepackage[utf8]{inputenc}
\usepackage[T1]{fontenc}
\usepackage{mathptmx}
\usepackage{url}

\begin{document}

\preprint{AIP/123-QED}

\title{Water diffusion in carbon nanotubes: interplay between confinement, surface deformation and temperature}

\author{Bruno H. S. Mendon\c{c}a}
\email{brunnohennrique13@gmail.com}
\affiliation{Instituto de F{\'i}sica, Universidade Federal do Rio Grande do Sul, Porto Alegre, RS, 91501-970, Brazil}

\author{Patricia Ternes}
\affiliation{School of Geography, University of Leeds, Leeds, LS2 9NL, UK}

\author{Evy Salcedo}
\affiliation{Coordenadoria Especial de F{\'i}sica, Qu{\'i}mica e Matem{\'a}tica, Universidade Federal de Santa Catarina, Ararangu{\'a}, SC, 88905-120, Brazil}

\author{Alan B. de Oliveira}
\affiliation{Departamento de F{\'i}sica, Universidade Federal de Ouro Preto, Ouro Preto, MG, 35400-000, Brazil}

\author{Marcia C. Barbosa}
\affiliation{Instituto de F{\'i}sica, Universidade Federal do Rio Grande do Sul, Porto Alegre, RS, 91501-970, Brazil}

\date{\today}

\begin{abstract}
In this article we investigate through molecular dynamics simulations the diffusion behavior of the TIP4P/2005 water when confined in pristine and deformed carbon nanotubes (armchair and zigzag). To analyze different diffusive mechanisms, the water temperature was varied from $210\leq T\leq 380$~K. The results of our simulations reveal that water present a non-Arrhenius to Arrhenius diffusion crossover. The confinement shifts the diffusion transition to higher temperatures when compared with the bulk system. In addition, for narrower nanotubes, water diffuses in a single line which leads to a mobility independent of the activation energy.
\end{abstract}

\maketitle

\section{Introduction}

Liquid water is as simple as it is strange. Interestingly, the fascination around water lies exactly in the connection between these two factors: liquid water behaves unexpectedly in many situations mostly because its molecule is so simple. Composed by only two different atoms, and being so small, it can move, rotate, vibrate, bond, and fit like now any other known liquid molecule. 

Not surprisingly, literature is vast when it comes to water and water related subjects.~\cite{CHA01,GUI02,FRA11,KUM06,COR09,COR11} The fact that life would not be possible if water was a normal liquid. Currently, there are seventy-four -- and counting -- unexpected properties associated to liquid water, which are known as water anomalies.~\cite{MAR} Some of those aforementioned anomalies are easily detected without any sophisticated apparatus. The canonical example is the bottle filled with liquid water, which cracks after some time in the freezer. The density of ice is smaller than the density of liquid water which means liquid water expands upon cooling, going against any reasonable thermodynamics argument. This is known as the water density anomaly. Other water anomalies are more involving. For example, its dynamics -- as measured by the diffusion coefficient -- is known to be anomalous as it increases under increasing pressures -- or, equivalently, under increasing densities --, for a certain range of temperatures. This is exactly the opposite behavior which is expected for a normal liquid under the same circumstances.~\cite{SPE76,NET01} 

Even more puzzling is the so-called water fragile-to-strong transition. Upon cooling, the dynamics of strong liquids slows down at a constant rate, while for fragile liquids such a rate increases with temperature drop. For a strong liquid, diffusion follows an Arrhenius, whereas a fragile liquid has a super-Arrhenius (or non-Arrhenius) behavior. Most liquids can be divided into these two categories, but water is, again, an exception. Water is fragile at ambient temperatures, while it appears to be strong upon supercooling. \cite{MAR16} Many explanations for such a fragile-to-strong transition have been proposed, being two of the most relevant the hypothesis proposed by Angel \emph{et al}.~\cite{ITO99} and Stanley \emph{et al}.~\cite{STA92}. While Angel's group associates the fragile-to-strong transition to the approximation of glass transition temperature, Stanley and collaborators ascribes it to the crossing of the Widom line due to its connection with the hypothetical liquid-liquid critical point of water.

It is worthy to mention, recent models on the footsteps of Stanley's idea have been treating water as being two liquids in one. Water is supposed to be composed by high- and low-density counterparts whose proportions vary with temperature, what would explain not only thermodynamic anomalies but also the dynamic ones. \cite{SHI18} This approach has gained attention of scientific community, despite with mixed receptivity.~\cite{SOP19,SHI18-2,FIT19} 

As we see, the normal for bulk water is to be abnormal. Then what would expect for such a liquid when it is constrained to highly confined environments?

Recently the thermodynamic and dynamic anomalies of water constrained to low dimensional geometries were revisited.~\cite{MOU05,STAN05,LIU06,KOH19} A side effect of confinement is that most of the anomalous behaviors are found in higher temperatures if compared with their bulk equivalents. In fact this allowed the experimental detection of fragile-to-strong crossover for confined water.~\cite{MOU05,LIU06} 

For studying confined water, carbon nanotubes (CNTs) offer an excellent laboratory. Their typical low diameter/length ratio makes them  nearly one-dimensional structures. Yet, their diameters can be as small as to only fit a single chain of water molecules, or big enough to approximate the confined water properties to bulk ones. This tuning feature is perfect for systematic investigations of the effects of confinement in liquid water. In addition, CNTs can be viewed as rolled graphene sheets. Graphene borders are different (being armchair or zigzag) and the way tubes are rolled results in different tube chiralities. This structural ingredient (which even affects the CNTs electronic structure) is also important when dealing with water confined in nanotubes.

The confinement of water inside nanotubes brings new phenomena which are not present in bulk water. For example, for water inside tubes with diameters above $6.0$~nm the diffusion coefficient is close to the bulk value. As the diameter decreases, simulations show that the diffusion decreases, reaching a minimum value for diameters of about $1.2$~nm.~\cite{YIN08,ALE08,NAN09} Then, water mobility raises for diameters smaller than $1.2$~nm.~\cite{FAR11,BOR12} Experiments suggest that at low diameters water dynamics breaks into a slow regime for the water molecules at the nanotube wall and a fast regime for the molecules close to the center.~\cite{HAS18,GKO20} This decoupling leads to a fragile-to-strong dynamic crossover observed experimentally  at very low temperatures.~\cite{MAM06}

The anomalous low diffusion observed for water confined in CNTs with diameters around $1.2$~nm is related to the ice-like structure assumed by water inside the channel. The increase of water mobility for diameters below this threshold is attributed to the smooth, inner hydrophobic surface of CNTs, which lubricates and speeds up a near-frictionless water transport.~\cite{FAL10,KAL02} 

For confined systems the main physics behind the diffusion includes the competition between two ingredients: local expansion or compression along chains of particles and fluid's confining surface mismatch.~\cite{BRA94,BRA96} The former is governed by local fluctuations in inter-molecule distances, thus imposing local tension or repulsion between particles. The later accounts for the imposed confining surface structure over the confined fluid molecules. In the particular case of confined water, these two mechanisms compete in order to optimize the hydrogen bond network. If the fluid-surface mismatch is large and local expansions are significant, the mobility is high. If the fluid-surface mismatch is small and if local densities won't bring additional tensions between particles, water molecules tend to accommodate themselves in local minima, decreasing the mobility. Thus, for the diffusion of confined water inside carbon nanotubes the important variables are -- but probably not limited to -- the nanotube diameter, degree of deformation (as measured by the eccentricity), chirality, and temperature.~\cite{TER17} 

The fluid-surface mismatch can be affected by nanotube deformation and chirality.~\cite{ANA01,WAN04,ANS13} In particular the chirality impacts the water dipole orientation due to a difference in the partial load distribution at the ends of the tubes.~\cite{CHA06,YIN08,YIN05,WAG17} In carbon nanotubes, defects are present either by irregularities in the crystal lattice or by the presence of adsorbed molecules.~\cite{OLI16} Structural irregularities lead to defect-induced mismatches between the tube and water chains.~\cite{BAO11,BHSM19,BRU20}

Even though the confined water dynamics have been widely studied, the understanding of how temperature affects the dynamic properties of water confined in CNT's with different topologies is still missing. In this work we analyze the impact of different chiralities, diameters, and temperatures in the diffusion of water confined in CTNs. We compare the mobility of water confined in perfect and in deformed nanotubes. We discuss the water mobility in the framework of local expansion or compression versus fluid-surface mismatch and hydrogen bond network. In particular we analyze what happens with the fragile-to-strong transition observed in bulk water. 

In the next section, we show the simulation details, in section III we present the results for water diffusion for various systems and section IV brings our conclusions.

\section{The model, simulation and methods}

Molecular dynamics (MD) simulations at constant number of particles, volume, and temperature were performed to analyze the diffusion coefficient of confined water. The water model used was the  TIP4P/2005~\cite{ABA05}, which presents good agreement with  experimental results, particularly for the diffusion coefficient.~\cite{VEG11} There are simpler models for water that still manage to reproduce some of their anomalies.~\cite{CAM03,CAM05}

\begin{figure}[!ht]
  \begin{center}
  \includegraphics[width=5.4in]{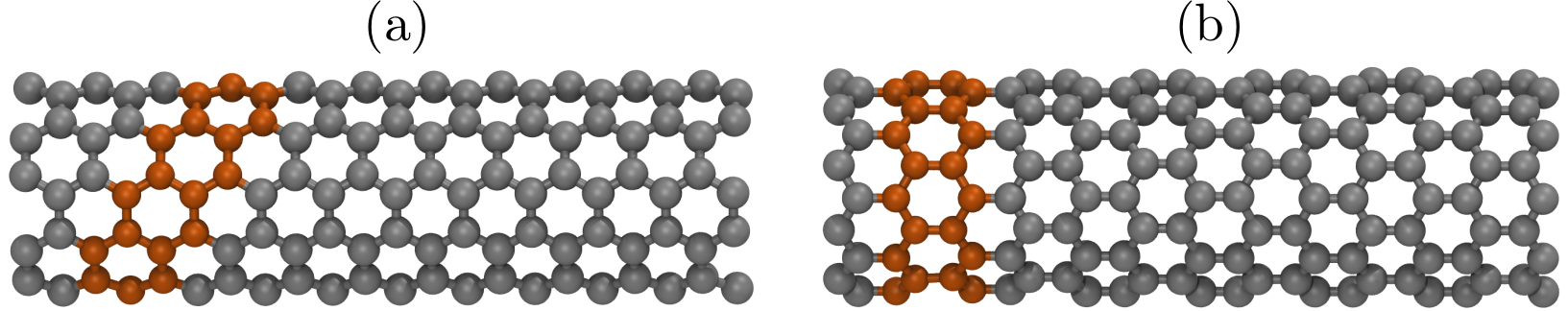}
  \end{center}
  \caption{Snapshot of (a) armchair and (b) zigzag carbon nanotubes. Highlighted atoms form  carbon rings in zigzag tubes and spirals in armchair ones.}     
  \label{fig_1}
\end{figure}

Water molecules were confined in CNTs with different diameters, chiralities, and deformations. Following the ($n,m$) notation for characterising the chirality of CNTs, we have used three armchair ($n=m$) nanotubes, namely, ($7,7$), ($9,9$), and ($12,12$), and three zigzag ($m=0$) nanotubes, which were ($12,0$), ($16,0$), and ($21,0$). Figure~\ref{fig_1} shows structural distinctions between armchair and zigzag nanotubes. The diameter of nanotubes may be given as a function of $n$ and $m$ indexes as follows

\begin{equation}
 d = \frac{\sqrt{3}}{\pi}a_{CC}\sqrt{n^2 + m^2 + nm},
\end{equation}

\noindent where $a_{CC}=1.42$~\AA\ is the C-C bond length (see Table~\ref{tab_CNT} for nanotubes dimensions).

\begin{figure}[!ht]
  \begin{center}
  \includegraphics[width=5.4in]{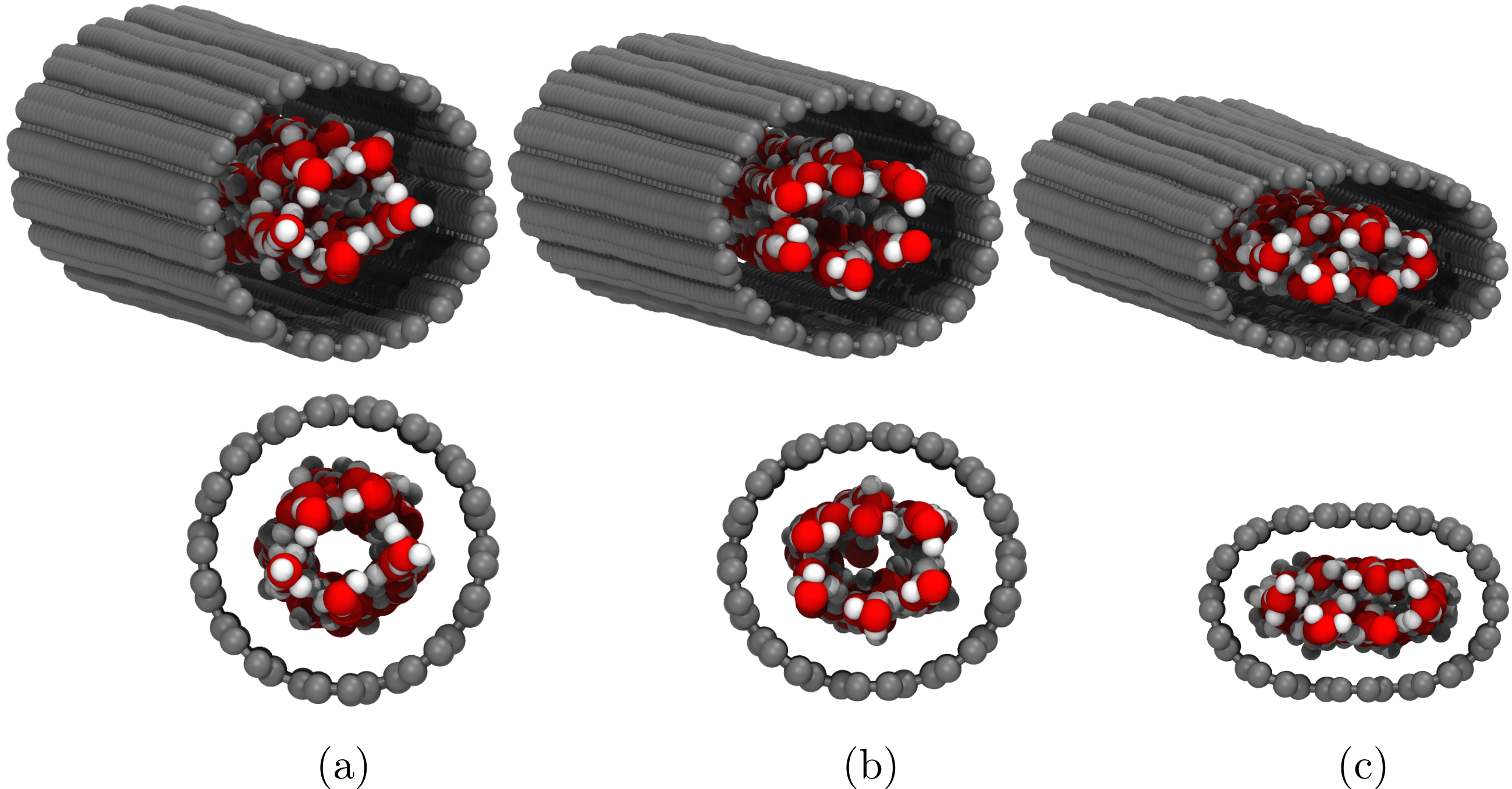}
  \end{center}
  \caption{Snapshots of a ($9,9$) nanotube with (a) $e=0.0$, (b) $e=0.4$ and (c) $e=0.8$.}     
  \label{fig_2}
\end{figure}

For investigating the effects of radial asymmetry in the diffusion of confined water we have uniformly deformed the nanotubes with different degrees. Here we define the deformation degree by the ellipse eccentricity 

\begin{equation}
  e= \sqrt{1 - \frac{a^{2}}{b^{2}}} \;,
\end{equation}

\noindent where $a$ is the  semi-minor axis and $b$ is the semi-major axis of a right section of the tube. Armchair and zigzag tubes were investigated for three degree of deformation, i.e., $e=0.0$ (perfect tube), $e=0.4$ and $e=0.8$ (see Figure~\ref{fig_2}). 

The carbon-water interaction was modeled through the Lennard-Jones (LJ) potential as follows.~\cite{LEN31} The carbon-oxygen energy $\epsilon_{CO}=0.11831$~kcal/mol and the effective carbon-oxygen diameter $\sigma_{CO}=3.28218$~\AA.~\cite{HUM01} The interaction between carbon and hydrogen was set to zero. Water density was determined considering the  excluded volume due the LJ interaction between carbon and oxygen atoms. Thus, the density is determined by $\rho= 4M/[\pi(d-\sigma_{CO})^2L_z]$, where $M$ is the total water mass into the tubes and $L_{z}$ is the nanotube length (see Table~\ref{tab_CNT}).~\cite{BOR12} Deformed nanotubes were simulated with the same length as well as the same total confined water mass that of the perfect equivalent nanotube.

\begin{table}[!ht]
\begin{center}
  \caption{Parameters for the simulated perfect carbon nanotubes ($e=0.0$) armchairs and zigzags.}
  \begin{tabular}{ c c c c c }
    \hline \hline
    CNT    $\quad$ & $\quad$ $d$ (nm) 	 & $\quad$ L$_{z}$ (nm) & $\quad$ $\rho$ (g/cm$^{3}$)\\ \hline
    (7,7)  $\quad$ & $\quad$ 0.95	 & $\quad$ 123.4        & $\quad$ 0.90               \\
    (12,0) $\quad$ & $\quad$ 0.94	 & $\quad$ 123.0        & $\quad$ 0.91               \\
    (9,9)  $\quad$ & $\quad$ 1.22	 & $\quad$ 50.5         & $\quad$ 0.92               \\
    (16,0) $\quad$ & $\quad$ 1.25	 & $\quad$ 50.5         & $\quad$ 0.80               \\
    (12,12)$\quad$ & $\quad$ 1.63	 & $\quad$ 22.5         & $\quad$ 0.94               \\
    (21,0) $\quad$ & $\quad$ 1.64	 & $\quad$ 22.9         & $\quad$ 0.86               
    \\ \hline \hline
  \end{tabular}
\label{tab_CNT}
\end{center}
\end{table}

Simulations were performed using Large-scale Atomic/Molecular Massively Parallel Simulator (LAMMPS) package.~\cite{LAMMPS} Periodic boundary conditions in the axial direction and a cutoff radius of  $12$~\AA\ in the interatomic potential were used. The water structure is constrained using the SHAKE algorithm~\cite{RYC77}, with tolerance of 1$\times10^{-4}$. Long-range Coulomb interactions were computed through the particle-particle particle-mesh (PPPM) method. In order to prevent real charges to interact  with their own images, we have set the simulations box dimensions perpendicular to the axial axis as $100$~nm.~\cite{OST17} The investigated temperature range  goes from $190$~K up to $380$~K, controlled by the Nos{\'e}-Hoover thermostat with a damping time of $100$~fs. In all simulations, nanotubes were kept rigid with zero center-of-mass velocity. This procedure has been employed in several similar simulations, having shown to be a very reasonable approximation when compared to the  case in which thermostat is applied throughout the whole system.~\cite{FAR11,HAN06,KOT04}

For maximizing computational efficiency different simulation times were used depending on the temperature range taken into consideration. For temperatures between $190$~K and $290$~K, a total time of $17$~ns were simulated, with the initial $10$~ns  used to equilibrate the system. Only the final $7$~ns were used to calculate system properties. For temperatures from $310$~K up to $380$~K, a total of $10$~ns were computed, with $5$~ns for equilibration purposes and the remaining 5 ns for production. The time step was $1$~fs in all runs, and properties were stored every $300$~fs. 

Due to the confining system geometry, the diffusion is minimal in the radial direction, therefore only the axial diffusion is considered. The axial diffusion coefficient is given by the one dimensional Einstein relation, namely

\begin{equation}
\label{eq:MSD}
D_z = \lim_{\tau \to \infty} \frac{1}{2}\frac{d}
{d\tau}\left\langle z^2(\tau)\right\rangle,
\end{equation}

\noindent where $\left\langle z^2(\tau) \right\rangle=\left\langle \left[z(\tau_0 - \tau) - z(\tau_0) \right]^2\right\rangle$ is the water mean square axial displacement, averaged over oxygen atoms.

On average, each run is composed of three sets of simulations with different initial thermal speed 
distribution. 

\section{Results and discussion}

\begin{figure}[!ht]
  \begin{center}
 \includegraphics[width=6.25in]{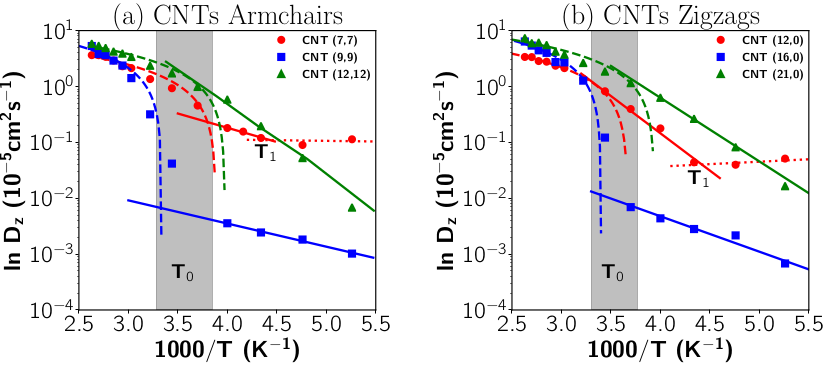}
  \end{center}
  \caption{Log of the diffusion coefficient versus inverse temperature for water inside (a) armchair and (b) zigzag nanotubes. $T_0$ stands for the region in which a non-Arrhenius to Arrhenius transition takes place, while $T_1$ is the location for (a) ($7,7$) and (b) ($12,0$) nanotubes to show a constant activation energy.}     
  \label{fig:perfect}
\end{figure}

First, we analyze the mobility of water inside perfect nanotubes for different temperatures and nanotube diameters. Figure~\ref{fig:perfect} shows the water diffusion versus inverse temperature for different armchair and zigzag nanotubes. For larger zigzag nanotubes the diffusion approaches the bulk value at high temperatures. For a constant temperature, it decreases with decreasing nanotube diameter as we would expect. 

Two processes can be observed under system cooling. First, a typical fragile-to-strong transition occurs at $T=T_0$. Secondly, the activation energy becomes constant for nanotubes with a critical diameter (between $0.9$ and $1.0$~nm) at $T=T_1$.  

For strong liquids the diffusion versus temperature dependence follows the Arrhenius equation,
\begin{eqnarray}
\label{eq:Arrhenius}
D = D_0 e^{-\Delta E/k_BT}
\end{eqnarray}
\noindent where $\Delta E$ is the activation energy for the diffusion and $D_0$ is a pre-exponential diffusion. 
As the system approaches $T_0$ the structural crossover from non-Arrhenius to the Arrhenius behavior occurs. This crossover temperature depends on the hydrogen bond network and, even though is lower than the temperature observed for bulk water, it does not depend strongly on the nanotube diameter or chirality.

For $T=T_1$, water molecules at  both ($12,0$) and ($7,7$) nanotubes show the peculiar behavior of water mobility  staying constant with the change of temperature. Water molecules inside those nanotubes assume a single line configuration, with formation all possible hydrogen bonds for a linear structure, and flowing as a single structure. The mobility at this regime seems  to be independent of wall structure and  changes in temperature do not contribute with any activation energy, and $\Delta E\approx 0$. The reason behind this non activation energy driven diffusion might be due  to the single-line arrangement of water inside the both ($12,0$) and ($7,7$) nanotubes which becomes almost on dimensional mobility.

\begin{figure}[!ht]
  \begin{center}
  \includegraphics[width=6.25in]{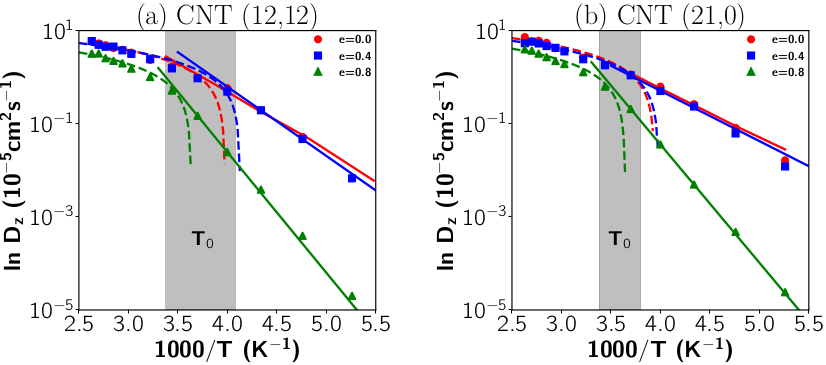}
  \end{center}
  \caption{Log of the diffusion coefficient versus inverse temperature for water inside (a) armchair ($12,12$) and (b) zigzag ($21,0$) nanotubes for different eccentricities. $T_0$ stands for the region in which a non-Arrhenius to Arrhenius transition takes place.}     
  \label{fig:deformed-12x12-21x0}
\end{figure}

Next, we check how these temperature regimes are affected by deformations in the nanotube, i.e., how the diffusion coefficient versus inverse temperature of water is affected by different degrees of the parameter $e$. For the larger diameters as illustrated in the Figure~\ref{fig:deformed-12x12-21x0} for both zigzag and armchair nanotubes the deformation decreases the water diffusion coefficient only for very large deformations. Even under deformation the non-Arrhenius to Arrhenius crossover is observed at a temperature $T_0$, and it is higher than the bulk value, which is between $222$~K and $200$~K.~\cite{MAR16} The low impact of $T_0$ confirms that the dynamic crossover is not very dependent of surface-mismatch effects.

\begin{figure}[!ht]
  \begin{center}
  \includegraphics[width=6.25in]{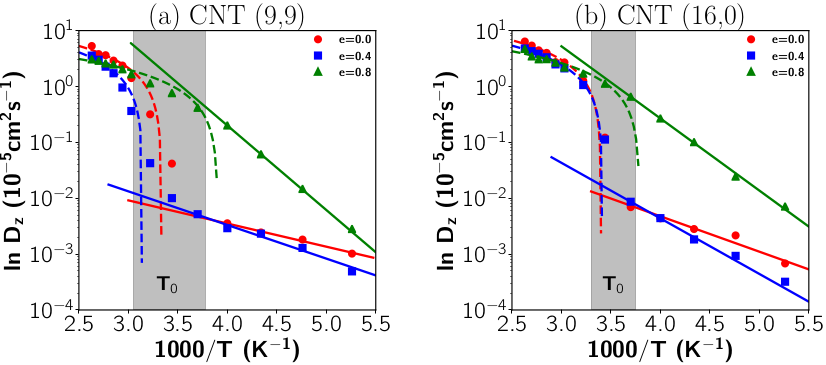}
  \end{center}
  \caption{Log of the diffusion coefficient versus inverse temperature for water inside (a) armchair ($9,9$) and (b) zigzag ($16,0$) nanotubes for different eccentricities. $T_0$ stands for the region in which a non-Arrhenius to Arrhenius transition takes place.}     
  \label{fig:deformed-9x9-16x0}
\end{figure}

Figure~\ref{fig:deformed-9x9-16x0} shows the diffusion coefficient for the perfect and deformed nanotubes versus inverse temperature for the ($16,0$) and ($9,9$) nanotubes. For all the three deformation cases there is a dynamic transition from non-Arrhenius to Arrhenius behavior for temperatures higher than the bulk value but that seems not to be affected by tube chirality.~\cite{MAR16} The non-Arrhenius behavior for the perfect and $e=0.4$, however, exhibit a quite unusual behavior with change in symmetry, as shown in Figure~\ref{fig:perfect}. This change is due to the ice-like structure formed for these systems not observed for the very deformed $e=0.8$. Since the very deformed systems are not ice-like, the water diffusion is enhanced by deformation.

\begin{figure}[!ht]
  \begin{center}
  \includegraphics[width=6.25in]{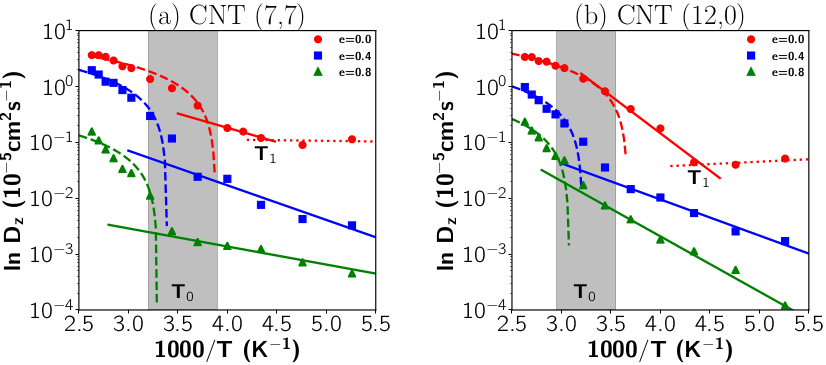}
  \end{center}
  \caption{Log of the diffusion coefficient versus inverse temperature for water inside (a) armchair ($7,7$) and (b) zigzag ($12,0$) nanotubes for different eccentricities. $T_0$ stands for the region in which a non-Arrhenius to Arrhenius transition takes place, while $T_1$ is the location for to show a constant activation energy.}     
  \label{fig:deformed-7x7-12x0}
\end{figure}

Finally, we show in the Figure~\ref{fig:deformed-7x7-12x0} the diffusion coefficient for water inside the ($12,0$) zigzag and ($7,7$) armchair nanotubes. For the perfect nanotube water shows three regimes, non-Arrhenius for $T>T_0$, Arrhenius for $T_1<T<T_0$ and a constant diffusion for $T<T_1$. As the nanotube is deformed the non-Arrhenius to Arrhenius regime persists but the constant diffusion observed for the perfect nanotube for $T<T_1$ disappears. This again is not surprising the single line behavior of water inside perfect both ($12,0$) and ($7,7$) nanotubes is disrupt by the deformation. Water moves in a non linear behavior. The deformed nanotube also present at low temperatures a lower mobility.

\section{Conclusions}

In this work we analyzed the water mobility under confinement in nanotubes with distinct diameters, chiralities and deformations. Each system was submitted to different temperatures ranging from $190$~K to $380$~K. For the perfect nanotube we observed two mechanisms for the diffusion.

At $T=T_0$, water presents a non-Arrhenius to Arrhenius crossover. Our results indicate that this behavior is defined by the hydrogen bond network, while confinement shifts this behavior to lower temperatures (when compared to bulk). This transition shows little dependence with the nature of the wall or degree of confinement. 

For the particular case of the ($12,0$) zigzag and ($7,7$) armchair nanotubes at $T<T_1$ another region in which the diffusion is independent of the temperature appears. Since this is present only for this specific diameter in which water forms a single line, this constant diffusion coefficient arises from the fluid-surface mismatch. 

The nanotube deformation produces two effects. For the water inside the ($16,0$) zigzag and ($9,9$) armchair nanotubes the deformation at low temperatures ``melts'' the ice-like structure present in the perfect nanotube. In the case of the water single line structure formed at the perfect ($12,0$) and ($7,7$) nanotubes, the deformation decreases the mobility and water do not present the constant diffusion for $T<T_1$.

The distinction between the water diffusion coefficient due to the change in the nanotube chirality only appears at low temperatures, where the armchair nanotube induces water to form an ice-like structure.

Due to the hydrophobic nature of carbon nanotubes, water molecules tend to avoid the surface. This fact plays a central role for water diffusion specially in narrow nanotubes. Aside of that, molecules also try to minimize energy by forming hydrogen bond networks. These two processes govern the mobility of confined water in a nontrivial way.

\begin{acknowledgments}
This work is partially supported by Brazilian science agencies CNPq~--~through INCT-Fcx~--, CAPES, FAPEMIG, Universidade Federal do Rio Grande do Sul (UFRGS), Universidade Federal de Ouro Preto (UFOP), and Centro Nacional de Processamento de Alto Desempenho (CENAPAD).
\end{acknowledgments}

\section*{Data Availability}

The data that support the findings of this study are available from the corresponding author upon reasonable request.


\bibliographystyle{apsrev4-1}
\bibliography{samp}
\end{document}